\documentclass[12pt]{article}

\usepackage[totalheight = 23cm, totalwidth = 17cm]{geometry}
\usepackage{amssymb,amsmath,amsfonts,amsbsy,graphicx}

\def\mathbi#1{\textbf{\em #1}}

\newcommand{\mpl}{m_{\rm Pl}}
\newcommand{\fnl}{f_{\rm NL}}
\newcommand{\calP}{{\cal P}}
\newcommand{\calR}{{\cal R}}
\newcommand{\leom}{{\left.\frac{\delta{L}}{\delta\calR}\right|_1}}

\begin{document}

\begin{titlepage}

\rightline{\footnotesize{APCTP-Pre2013-015}} \vspace{-0.2cm}
\rightline{\footnotesize{MAD-TH-13-08}} \vspace{-0.2cm}

\begin{center}

\vskip 1.0cm

\Large{Correlating correlation functions of  primordial perturbations}

\vskip 1.0cm

\large{
Jinn-Ouk Gong$^{a,b}$,
\hspace{0.2cm} 
Koenraad Schalm$^{c,d}$
\hspace{0.2cm} and \hspace{0.2cm}
Gary Shiu$^{e,f}$
}

\vskip 0.5cm

\small{\it 
$^{a}$Asia Pacific Center for Theoretical Physics, Pohang 790-784, Korea 
\\
$^{b}$Department of Physics, Postech, Pohang 790-784, Korea
\\
$^{c}$Department of Physics, Harvard University, Cambridge, MA 02138, USA
\\
$^{d}$Instituut-Lorentz for Theoretical Physics, Universiteit Leiden, 2333 CA Leiden, The Netherlands
\\
$^{e}$Department of Physics, University of Wisconsin, Madison, WI 53706, USA
\\
$^{f}$Center for Fundamental Physics \& Institute for Advanced Study,
\\ Hong Kong University of Science and Technology,
Clear Water Bay, Hong Kong
}

\vskip 1.2cm

\end{center}

\begin{abstract}

We explore the correlations between correlation functions of the primordial curvature perturbation produced during inflation. We find that for general single field inflation, other than the source terms which depend on the model details, higher order correlation functions are characterized by the power spectrum, its spectral index and running. The correlation between the bispectrum and power spectrum is presented as an explicit example of our systematic approach.

\end{abstract}

\end{titlepage}

\setcounter{page}{0}
\newpage
\setcounter{page}{1}

\section{Introduction}

Recent progress in understanding effective field theories for time-dependent backgrounds has significantly broadened the scope of inflation and its predictions~\cite{Cheung:2007st}. Heavy field effects which are justifiably described in effective field theories are shown to be capable of leaving observable imprints in a cosmological background by reducing the effective speed of sound for the curvature perturbation~\cite{heavy,Achucarro:2012sm}. Inflationary dynamics involving heavy degrees of freedom can lead to oscillatory features in the power spectrum and a resonantly enhanced bispectrum~\cite{resonantbi,Achucarro:2012fd} (see also~\cite{chen}). The outliers in the power spectrum of the cosmic microwave background~\cite{Ade:2013uln} and hints of primordial oscillatory signals in the bispectrum~\cite{Ade:2013ydc} could well be echoing the effects of heavy fields during inflation.

More generally, features that leave transient signals in the observable correlation functions can be generated by various mechanisms, such as a kink in the inflaton potential~\cite{Starobinsky:1992ts}. Once generated, the effects permeate any correlation function through the mode function solution of the curvature perturbation. Thus,  the correlation functions of the primordial perturbation are all correlated and this offers a unique opportunity to detect higher order correlation functions. Given the null detection of the bispectrum from PLANCK~\cite{Ade:2013ydc}, detecting signals correlated in a certain manner is much more probable than blindly looking for them.

In this article, we study how to correlate the correlation functions of the primordial curvature perturbation by using  the general slow-roll formalism (GSR)~\cite{gsr}. Explicitly, we find that, other than source terms, higher order correlation functions can be characterized by the power spectrum, its spectral index and running for general single field inflation, regardless of whether there are features or not. Thus our formalism has a wide range of applicability to inflationary scenarios. While the consistency condition of~\cite{Maldacena:2002vr} relates the three-point function to the two-point one in the squeezed limit\footnote{The consistency condition of~\cite{Maldacena:2002vr}, which was generalized to higher order functions in~\cite{Huang:2006eha}, was used in~\cite{Jackson:2013vka} to derive a theoretical upper bound on the frequency of oscillations in the scalar perturbation power spectrum of single field inflation.}, the correlations we found are not limited to a special kinematic limit. We present the bispectrum\footnote{In~\cite{Adshead:2011bw} the bispectrum was computed based on GSR, emphasizing the case with features in the inflaton potential. In this article we consider more general possibilities, and more importantly, we discuss the bispectrum as a non-trivial example to show the explicit correlation to the power spectrum.} as an example and discuss two simple cases where correlated features in the power spectrum and bispectrum can arise.

\section{General slow-roll formalism}

We first extend GSR to general single field theory with the matter Lagrangian $P(X\equiv-\partial^\mu\phi\partial_\mu\phi/2,\phi)$ in a form appropriate for our purpose. At second order, the general effective action for the comoving curvature perturbation $\calR$ is
\begin{equation}
\label{S2}
S_2 = \int dtd^3 \mathbi{x} ~a^3\mpl^2\epsilon \left[ \frac{\dot\calR^2}{c_s^2} - \frac{(\nabla\calR)^2}{a^2} \right] \, ,
\end{equation}
where $\epsilon \equiv -\dot{H}/H^2$ and $c_s$ is the speed of sound. Introducing new variables~\cite{Baumann:2011dt}  $z^2 \equiv 2a^2\mpl^2\epsilon/c_s$, $y \equiv \sqrt{2k}z\calR_k$, $dx\equiv-kd\tau\equiv-kc_sdt/a$ and $f \equiv 2\pi xz/k = f(\log\tau)$, we find the mode equation which follows from this action given by
\begin{equation}\label{eom2}
\frac{d^2y}{dx^2} + \left( 1 - \frac{2}{x^2} \right) y = \frac{1}{x^2} \frac{f''-3f'}{f} y \equiv \frac{g(\log\tau)}{x^2} y \, ,
\end{equation}
where $f' = df/d\log\tau$. The homogeneous solution which satisfies the boundary condition to reproduce the usual Minkowski mode function as $-k\tau\to\infty$ is the pure de Sitter mode function,  $y_0(x)=(1+i/x)e^{ix}$. Then, with $g$ being a small perturbation, we can write the perturbative solution of $y(x)$ using the Green's function \cite{Gong:2001he}:
\begin{align}\label{GSRsol}
y(x) = & y_0(x) + \frac{i}{2} \int_x^\infty \frac{du}{u^2} g(\log{u}) \Big[ y_0^*(u)y_0(x) - y_0(u)y_0^*(x) \Big] y(u) \, .
\end{align}
Then, up to the first order corrections, we have the power spectrum~\cite{gsr}
\begin{equation}\label{GSRpower}
\calP_\calR(k) = \lim_{\tau\to0} \frac{1}{f^2}  \left[ 1-\frac{2}{3}\tau^3 \int_\tau^\infty \frac{d\tau'}{{\tau'}^4} g(\log\tau') + \frac{2}{3} \int_\tau^\infty \frac{d\tau'}{\tau'}g(\log\tau')W(k\tau') \right] \, ,
\end{equation}
where $W(x) \equiv -3\Re[y_0(x)]\Im[y_0(x)]/x$, and we can define another window function $X(x) \equiv 3\left\{ \Re[y_0(x)] \right\}^2/x$ [see (\ref{GSRbi})]. This formula is valid for models where the usual slow-roll approximation is not effective because of features in the inflaton potential~\cite{Gong:2005jr}.

Given that we have expressed the power spectrum purely in terms of the GSR fundamental function $f=f(\log\tau)$ and its first and second derivatives, we can invert (\ref{GSRpower}) to write $f$ in terms of $\calP_\calR$ \cite{inverse}:
\begin{equation}
\begin{split}
\log\left( \frac{1}{f^2} \right) & = \int_0^\infty \frac{dk}{k} m(k\tau)\log\calP_\calR(k) \, ,
\\
m(x) & = \frac{2}{\pi} \left[ \frac{1}{x} - \frac{\cos(2x)}{x} - \sin(2x) \right] \, .
\end{split}
\end{equation}
Hence, $f'/f$ as well as $g$, the source of the power spectrum, can be written in terms of $\calP_\calR$ and its first two derivatives, i.e. the spectral index and the running:
\begin{align}
\label{inverse1}
\frac{f'}{f} & = \frac{1}{2} \int_0^\infty \frac{dk}{k} m(k\tau) \frac{d\log\calP_\calR}{d\log k} \, ,
\\
\label{inverse2}
g  &= -\frac{1}{2} \int_0^\infty \frac{dk}{k} m(k\tau) \left[ \frac{d^2\log\calP_\calR}{d\log k^2} + 3\frac{d\log\calP_\calR}{d\log k} \right] + \left[ \frac{1}{2} \int_0^\infty \frac{dk}{k} m(k\tau) \frac{d\log\calP_\calR}{d\log k} \right]^2 \, .
\end{align}
In summary, given the power spectrum, we can construct the GSR fundamental function $f$ and thus the mode function solution (\ref{GSRsol}). This in turn enables us to explicitly correlate other higher order correlation functions with the power spectrum.

\section{Sample correlation: bispectrum}

To be more specific, we now compute the bispectrum and explicitly correlate it to the power spectrum. Starting from the cubic order action~\cite{Maldacena:2002vr,Chen:2006nt} (see also \cite{Seery:2005wm}), an immediate difficulty we face when we apply GSR is that with many time derivatives the formulae become highly non-trivial. This is because the full GSR solution $y(x)$ is a perturbative series and the number of  terms we should incorporate increases rapidly. Thus, we first need to reduce the number of time derivatives to save our labour. For this, we make use of the linear equation of motion,
\begin{equation}
\leom \equiv \frac{a^3\epsilon}{c_s^2} \left\{ \ddot\calR + \left[ \frac{c_s^2}{a^2\epsilon}\frac{d}{dt} \left( \frac{a^2\epsilon}{c_s^2} \right) + H \right] \dot\calR - \frac{c_s^2}{a^2}\Delta\calR \right\} \, ,
\end{equation}
where $\eta \equiv \dot\epsilon/(H\epsilon)$ and $s \equiv \dot{c}_s/(Hc_s)$. Denoting schematically the coefficient of $\dot\calR$ inside the square brackets as $C = H\left(3+\eta-2s\right)$, the term $\dot\calR^3$ in the cubic action can be written with one time derivative on $\calR^3$ as
\begin{align}
\int A\dot\calR^3 &= \int \left[ \frac{\ddot{A}-3\dot{A}C-2A\dot{C}+2AC^2}{2} \frac{d}{dt} \left( \frac{\calR^3}{3} \right) + \cdots + \leom \frac{c_s^2}{a^3\epsilon} \left( \frac{\dot{A}-2AC}{2}\calR^2 + \cdots \right) \right] \, ,
\end{align}
where $A$ denotes schematically the coefficient of $\dot\calR^3$, and we have omitted the spatial gradient terms which are suppressed on large scales. Thus we see  that essentially we can reduce $\dot\calR^3$ term to $d\left(\calR^3\right)/dt$ with a field redefinition which includes more terms multiplied by the linear equation $\delta{L}/\delta\calR|_1$. We can proceed similarly for the $\dot\calR^2\calR$ term, and we find, with $B$ denoting the coefficient schematically,
\begin{align}
\int B\dot\calR^2\calR &= \int \left[ \frac{-\dot{B}+BC}{2} \frac{d}{dt} \left( \frac{\calR^3}{3} \right) + \cdots + \leom \frac{c_s^2}{a^3\epsilon}\frac{-B}{2}\calR^2 \right] \, .
\end{align}
Explicitly writing out the coefficients, the cubic order action from the matter sector, which is less suppressed than that from the gravitational sector and thus more relevant for observations, is given by
\begin{align}
S_3 & = \int d\tau d^3 \mathbi{x} ~ \frac{a^2\mpl^2\epsilon}{-c_s\tau} \left[ -\frac{c_saH\tau}{\mpl^2H^2\epsilon} \left\{ \left( 3s - \epsilon s + \eta s - 3us - \frac{\dot{s}}{H} - \frac{\epsilon\eta u}{2} + \epsilon us - \eta us  + \frac{u}{2}\frac{\dot\eta}{H} \right)\Sigma \right.\right.
\nonumber\\
& \hspace{3.5cm} + \left[ 6\epsilon-3\eta+6s - 6\epsilon^2 - 6\epsilon^2 + 4\epsilon\eta - 2\eta^2 - 14\epsilon s + 8\eta s - 8s^2 + 2\frac{\dot\eta}{H} - 4\frac{\dot{s}}{H} \right.
\nonumber\\
& \hspace{4cm} \left.\left. + \left( 3 - 5\epsilon + 3\eta - 6s \right)l - \left( l^2 + \frac{\dot{l}}{H} \right) \right] \lambda \right\}
\nonumber\\
& \hspace{3.5cm} \left. + c_saH\tau \left( 3s+\frac{\epsilon\eta}{2} - \epsilon s - 6us \right) - \frac{1}{2} \frac{d}{d\log\tau} \left( \frac{\eta}{c_s^2} \right) \right] \frac{d}{d\tau} \left( \frac{\calR^3}{3} \right) \, ,
\end{align}
where $u\equiv1-1/c_s^2$, $\Sigma \equiv \epsilon\mpl^2H^2/c_s^2$, $\lambda \equiv X^2P_{XX} + 2X^3P_{XXX}/3$ and $l \equiv \dot\lambda/(H\lambda)$.

The ``source function'' $g_B$ for the bispectrum is obtained by multiplying  the coefficient of $d\left(\calR^3\right)/d\tau$ by $-c_s\tau/\left( a^2\mpl^2\epsilon f \right)$. In particular, if we are interested in the case where a heavy degree of freedom is responsible for a non-unity $c_s$~\cite{heavy}, we find simply that
\begin{align}
g_B(\log\tau) = \frac{1}{f} &\left[ c_saH\tau \left( 3s + \frac{\epsilon\eta}{2} - \epsilon s - 3us - 2s^2 \right) + \frac{1}{2} \frac{d}{d\log\tau} \left( \frac{\eta}{c_s^2} \right) \right] \, .
\end{align}
Then, the bispectrum up to first order corrections is found to be~\cite{Adshead:2011bw}
\begin{align}\label{GSRbi}
B_\calR(k_1,k_2,k_3) = & \frac{(2\pi)^4}{4} \frac{\sqrt{\calP_\calR(k_1)}}{k_1^2} \frac{\sqrt{\calP_\calR(k_2)}}{k_2^2} \frac{\sqrt{\calP_\calR(k_3)}}{k_3^2} \int_0^\infty \frac{d\tau}{\tau} g_B(\log\tau)
\nonumber\\
& \times \left\{ \left( d_\tau-3\frac{f'}{f} \right) W_B + \frac{1}{3}d_\tau \left( X_B + X_{B3} \right) \int_0^\infty \frac{d\tilde\tau}{\tilde\tau} g(\log\tilde\tau) X(k_3\tilde\tau) + \text{2 perm} \right.
\nonumber\\
& \hspace{0.5cm} - \frac{1}{3}d_\tau W_{B3} \int_\tau^\infty \frac{d\tilde\tau}{\tilde\tau} g(\log\tilde\tau) W(k_3\tilde\tau) - \frac{1}{3}d_\tau X_{B3} \int_0^\tau \frac{d\tilde\tau}{\tilde\tau} g(\log\tilde\tau) X(k_3\tilde\tau) + \text{2 perm}
\nonumber\\
& \left. \hspace{0.5cm} - \frac{1}{2}d_\tau \left( X_B+X_{B3} \right) \int_\tau^\infty \frac{d\tilde\tau}{\tilde\tau} g(\log\tilde\tau) \left( \frac{1}{k_3\tilde\tau} + \frac{1}{k_3^3\tilde\tau^3} \right) + \text{2 perm} \right\} \, ,
\end{align}
where $f'/f$ and $g$ are given by (\ref{inverse1}) and (\ref{inverse2}) respectively in terms of $\calP_\calR$ and its derivatives, $d_\tau \equiv d/d\log\tau + 3$ and additional window functions are constructed from $y_0$ as
\begin{equation}
\begin{split}
y_0(k_1\tau)y_0(k_2\tau)y_0(k_3\tau) & \equiv W_B(k_1,k_2,k_3;\tau) + iX_B(k_1,k_2,k_3;\tau) \, ,
\\
y_0(k_1\tau)y_0(k_2\tau)y_0^*(k_3\tau) & = W_B(k_1,k_2,-k_3;\tau) + iX_B(k_1,k_2,-k_3;\tau) \equiv W_{B3}+iX_{B3} \, .
\end{split}
\end{equation}
In (\ref{GSRbi}), the implicit dependence of $\tau$ on $c_s$ does not matter practically, since we can make all the inner integrations with respect to $\tilde\tau$ from 0 to $\infty$ by incorporating appropriate step functions. Then the information on the speed of sound is encoded in $f$, and in turn in $\calP_\calR$.

Before we proceed, we turn to the effective field theory point of view~\cite{Cheung:2007st}. In that perspective, the effective action is specified by a set of mass scales $M_n^4$, which are essentially the expansion parameters of the effective field theory and are in principle independent couplings. Up to cubic order, the action is completely specified by $M_2^4$ and $M_3^4$. As can be read from (\ref{S2}), $c_s$ is determined at second order, viz. by $M_2^4$, but $M_3^4$ is to be specified independently~\cite{Gong:2013sma}. This appears in (\ref{GSRbi}) as the bispectrum source $g_B$, which in general is independent from the power spectrum source $g$. Nevertheless, we expect that $M_3$ is related to $M_2$ with the precise form being dependent on specific models. For example, $M_3^4/M_2^4 = 3\left(1-c_s^{-2}\right)/2$ for DBI inflation~\cite{Alishahiha:2004eh} and $3\left(1-c_s^{-2}\right)/4$ when a single heavy degree of freedom is integrated out~\cite{Achucarro:2012sm}. The source $g_B$ can be explicitly connected to $\calP_\calR$ for specific cases~\cite{Achucarro:2012fd}.

\begin{figure}[t]
 \begin{center}
  \includegraphics[width=15cm, trim=3.5cm 16.5cm 3.5cm 4cm]{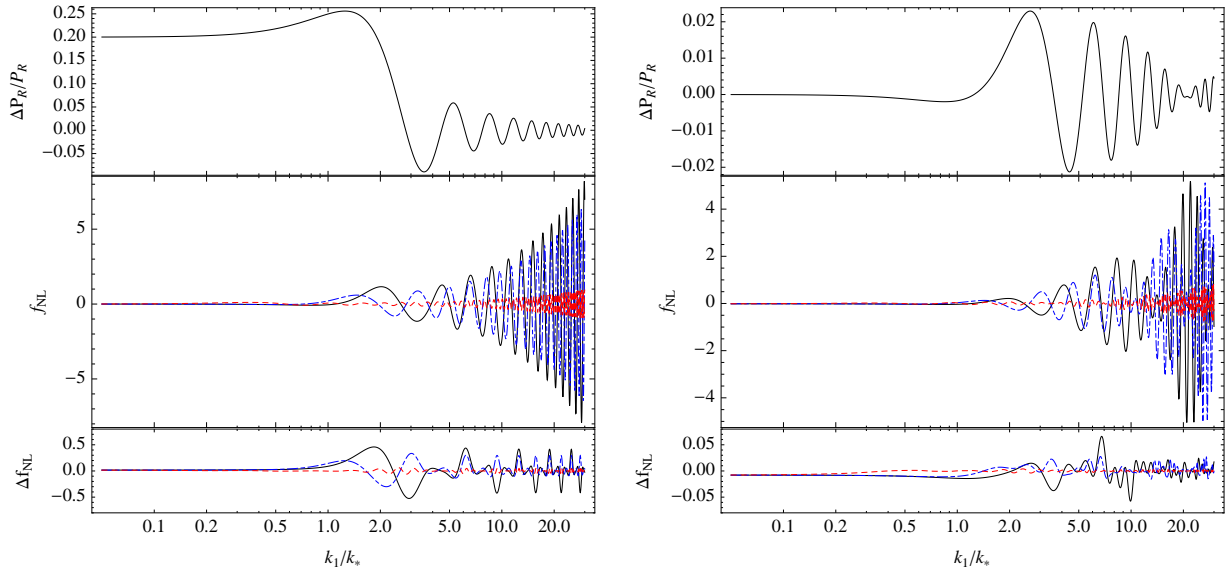}
 \end{center}
 \caption{Plot of $\Delta\calP_\calR/\calP_\calR$ (top panel) and $\fnl$ projected on equilateral (solid), folded (dot-dashed) and squeezed (dashed) configurations (middle panel). We also present the difference $\Delta\fnl = \fnl\text{(direct)} - \fnl\text{(inverse)}$ (bottom panel). In the left panel we show the linear slope change with $\Delta{A}/A=0.1$, and in the right panel the case is presented where $c_s^{-2}=1.1$ for limited $e$-folds $\Delta{N}=0.15$, otherwise $c_s^{-2}=1$.}
 \label{fig:examples}
\end{figure}

We now present two simple illustrative examples. We specifically choose examples in which features arise to exhibit the generality of our approach vis-\`a-vis e.g. effective field theory approaches. First we consider an inflation model~\cite{Starobinsky:1992ts} where the slope of the linear potential changes abruptly from  $-A$ to $-A-\Delta{A}$ at $\phi=\phi_0$. The usual slow-roll condition is violated at $\phi_0$ since $V''$ experiences a sharp change, which causes oscillatory features. If $|\Delta{A}/A|\ll1$ so that the approximate scale invariance of $\calP_\calR$ is maintained, we can write the bispectrum source as $g_B=-(f'/f)'/f$. In the left panel of Figure~\ref{fig:examples} we show the results with $\Delta{A}/A=0.1$ so that the slope becomes steeper with time. Then the power spectrum is suppressed for $\phi>\phi_0$, as shown in the plot. To estimate the shape and magnitude of the bispectrum, we may define a dimensionless shape function as
\begin{equation}
\fnl(k_1,k_2,k_3) \equiv \frac{10}{3} \frac{k_1k_2k_3}{k_1^3+k_2^3+k_3^3} \frac{(k_1k_2k_3)^2B_\calR}{(2\pi)^4\Delta_\calR^4} \, ,
\end{equation}
where $\Delta_\calR^2 = 2.215 \times 10^{-9}$ is the observed amplitude of the power spectrum~\cite{Ade:2013zuv}. In the lower left panel shown are $\fnl(k_1,k_2,k_3)$ projected on equilateral ($k_2/k_1=k_3/k_1=1$, solid line), folded ($k_2/k_1=1$ and $k_3/k_1=2$, dot-dashed line) and squeezed ($k_2/k_1=1$ and $k_3/k_1\ll1$, dashed line) limits, and the difference of $\fnl$ computed by direct calculations and by inverse formula, $\Delta\fnl = \fnl\text{(direct)} - \fnl\text{(inverse)}$. Likewise, in the right panel of Figure~\ref{fig:examples} we show the case where $c_s^{-2}$ exhibits a top-hat change~\cite{Nakashima:2010sa} from 1 to 1.1 for limited $e$-folds 0.15.

\section{Conclusions}

To conclude, we have presented a systematic approach for correlating the correlation functions of the primordial curvature perturbation produced during inflation. We have shown that in general, other than the source terms, any higher order correlation functions are characterized by the power spectrum, its spectral index and running. This approach should be useful in the search for higher order correlation functions of primordial perturbations.

\subsection*{Acknowledgements}

We thank Ana Achucarro, Peter Adshead, Gonzalo Palma, Subodh Patil, David Seery, Leonardo Senatore and Shuichiro Yokoyama for helpful discussions.
JG thanks the Aspen Center for Physics for hospitality, supported in part by the National Science Foundation under Grant No. 1066293, where part of this work was carried out.
JG and KS are very grateful to the Hong Kong Institute for Advanced Study where this project started. 
GS would like to thank Universiteit van Amsterdam for hospitality while he was visiting the Institute for Theoretical Physics as the Johannes Diderik van der Waals Chair.
He further thanks the Instituto de F\'isica Te\'orica UAM-CSIC for hospitality while this work was completed.
JG and GS thank the Lorentz Center, Universiteit Leiden for hospitality where this work was under progress.
JG acknowledges the Max-Planck-Gesellschaft, the Korea Ministry of Education, Science and Technology, Gyeongsangbuk-Do and Pohang City for the support of the Independent Junior Research Group at the Asia Pacific Center for Theoretical Physics. JG is also supported by a Starting Grant through the Basic Science Research Program of the National Research Foundation of Korea (2013R1A1A1006701). 
KS is supported in part by a VICI grant of the
Netherlands Organization for Scientific Research (NWO), by the
Netherlands Organization for Scientific Reseach/Ministry of Science
and Education (NWO/OCW) and by
the Foundation for Research into Fundamental Matter (FOM).
GS is supported in part by a DOE grant under Contract No. DE-FG-02-95ER40896.

\end{document}